# Volatile optical bistability enabled by mechanical nonlinearity

Dimitrios Papas[1], Jun-Yu Ou[1], Eric Plum[1] and Nikolay I. Zheludev[1,2]

[1] *Optoelectronics Research Centre and Centre for Photonic Metamaterials, University of Southampton, Highfield, Southampton SO17 1BJ, United Kingdom*

[2] *Centre for Disruptive Photonic Technologies, SPMS, TPI, Nanyang Technological University, Singapore 637371, Singapore*

**Optical devices with metastable states controlled with light (optical flip-flops) are needed in data storage, signal processing and displays. Although non-volatile optical memory relying on structural phase transitions in chalcogenide glasses has been widely used for optical data storage, beyond that, weak optical nonlinearities have hindered the development of low-power bistable devices. Here we report on a new type of volatile optical bistability in a resonant hybrid nano-optomechanical device, comprising of a pair of anchored nanowires decorated with plasmonic metamolecules. The nonlinearity resides in the mechanical properties of the nanowires and is transduced to its optical response by reconfiguring the plasmonic metamolecules. Such a system can be driven to a bistable response by acoustic signals modulated at the natural mechanical resonance of the nanowire. The memory of such a device is volatile and can be erased by removing the acoustic signal but in its presence, it can be switched between bistable optical states with microwatts of optical power. We argue that the demonstration of hybrid nano-optomechanical bistability opens new opportunities to develop practical low-power bistable devices.**

## Introduction

Bistability manifests as two stable output states of a nonlinear system for the same input. Optical bistability is needed for routing and memory in photonic circuits and computing[1]. Generally, an optically bistable system consists of a nonlinear medium with an external[2, 3] or internal[4] feedback mechanism and exhibits hysteresis as a function of light intensity. The earliest examples consist of sodium vapour[2], GaAs[3] or another nonlinear material inside an optical cavity, where nonlinearity of the complex refractive index (Kerr effect and saturable absorption) leads to dispersive and absorptive bistability. Since then, optical bistability has been reported in photonic crystal microresonators[5-7], ring resonators[8-10] and lasers[11], where high-Q resonators enable low-power bistability at the cost of low optical bandwidth. In metamaterials, bistability has been limited to microwave frequencies[12-15], where large nonlinearities are available. Weakness of optical nonlinearities has prevented practical, broadband, low-power solutions for bistability in the optical spectral range. In contrast, strong mechanical nonlinearities are easily accessible. In micro- and nanoelectromechanical systems (MEMS and NEMS), bistable cantilevers and beams can provide mechanical data storage and computing[16-25]. For instance, two buckling states[26] of nanostructures can serve as non-volatile mechanical memory states [21-23] or quantum qubits[27], controlled by optical, electrostatic or magnetic forces. Here, we demonstrate how mechanical nonlinearity can be exploited to achieve low-power, bistable states with distinctively different optical properties in a hybrid nano-optomechanical system. The bistable structure under study consists of metamaterial nanowires, the fundamental building blocks of a nanomechanical metamaterial[28] that combine resonant and nonlinear mechanical properties with resonant optical properties (Fig. 1). The bistable states can be controlled both by microwatts of light power and dissipate picowatts of mechanical (acoustic) power, opening up an opportunity to develop low-power bistable devices based on optomechanical bistability of nanomechanical photonic metamaterials.

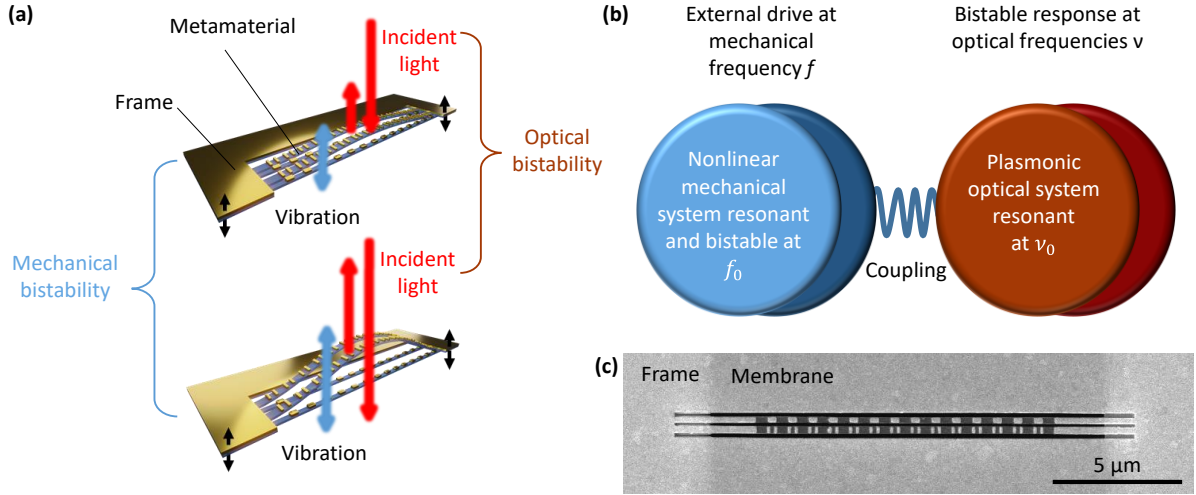

Fig. 1. Optomechanical bistability. (a) Mechanical nonlinearity of doubly-clamped metamaterial nanowires results in mechanical bistability – different amplitudes of mechanical oscillation under the same driving conditions. Transduction of this oscillation to modulation of light results in bistability of the structure's optical properties. (b) The bistability of the coupled system occurs at mechanical resonances at frequencies $f_0$ and may be controlled either by the acoustic driving signal at frequency $f$ or by incident light at frequency $\nu$. (c) SEM image of the bistable metamaterial nanowires consisting of gold plasmonic nanorods supported by silicon nitride beams.

## Results

*Origin of the bistability*

Resonant structures driven in the nonlinear regime can exhibit a bistable response[29]. A well-known example is a doubly clamped beam described by the Duffing equation of motion of a nonlinear oscillator[30]

$$m\ddot{z} + \gamma\dot{z} + m\Omega_0^2 z + \alpha z^3 = F_0 cos(\Omega t) \qquad (1)$$

with mass $m$, damping $\gamma$, $\Omega_0 = 2\pi f_0$ with resonance frequency $f_0$, cubic nonlinearity $\alpha$, oscillation amplitude $z$, driving force $F_0$ and $\Omega = 2\pi f$ with driving frequency $f$. In the case of a nanowire, the mechanical nonlinearity arises from increased stress along the beam at sufficiently high displacement amplitudes, resulting in frequency pulling and the formation of a bistable region around the oscillator's resonance frequency. The two stable states can serve as dynamic mechanical memory[17-20] and be accessed with increasing/decreasing driving frequency $f$ such that it approaches the structure's resonance frequency $f_0$ from below/above.

The bistable beam (nanowire) may also be controlled optically. Optical power $P$ incident on the beam with absorption coefficient $A$ and thermal resistance $R_{th}$ causes an average temperature increase $\Delta T = PAR_{th}$. This reduces the tensile stress $\sigma$ along the beam with initial stress $\sigma_0$, Young's modulus $E$ and thermal expansion coefficient $\alpha$ according to $\sigma = \sigma_0 - \alpha E \Delta T$. The mechanical resonance frequency of a tensile stressed doubly-clamped beam of thickness $h$, length $L$ and density $\rho$ is[31, 32]

$$f_0(\sigma) = 1.03 \frac{h}{L^2}\sqrt{\frac{E}{\rho}}\sqrt{1 + \frac{\sigma L^2}{3.4 E t^2}} \ . \qquad (2)$$

Therefore, the optical power incident on a lossy doubly-clamped beam controls its mechanical resonance frequency. We will demonstrate that this enables optical switching between the two bistable mechanical states at a fixed mechanical driving frequency $f$, resulting in mechanically-enabled optical bistability. In this case, the two stable states are accessed by increasing/decreasing optical power such that the structure's decreasing/increasing resonance frequency $f_0$ approaches the driving frequency $f$ from above/below.

*Metamaterial nanowires*

The optomechanical bistability may be expected in nanomechanical photonic metamaterials, which modulate light through reconfiguration of metamolecules[33] and relative displacement of metamaterial components or their arrays.[28]

We demonstrate memory effects utilizing a plasmonic nanomechanical system previously shown to exhibit light-induced modulation of its optical properties[34]. The structure consists of a pair of doubly-clamped silicon nitride nanowires (beams) of 15.8 µm length with fundamental mechanical resonance frequencies $f_0$ in the MHz range. The nanowires support an array of Π-shaped plasmonic meta-molecules (optical resonators), each consisting of three gold nanorods (Fig. 1c). Different parts of the plasmonic metamolecule are supported on different nanowires, to allow for mutual displacement between components of the metamolecule, making the optical response sensitive to relative nanowire displacement. The plasmonic structure is resonant at frequency *v*=230 THz (wavelength of 1310 nm). At this resonance, incident light excites a "bright" dipole mode in the larger nanorod, which couples to an anti-symmetric "dark" mode in the pair of smaller nanorods. The metamaterial nanowires were fabricated by focused ion beam milling from a 50 nm thick, low stress (<250 MPa) silicon nitride membrane coated with 50 nm of thermally evaporated gold. The ends of the nanowires are anchored to a 200 µm thick silicon frame. This provides fixed boundary conditions for the two nanowires, isolating the nanowires from natural modes of the rest of the membrane and yielding increased tensile stress in the nanowires upon displacement.

The experiments were performed in a vacuum cell at 4 x 10$^{-3}$ mbar pressure to avoid air-damping of mechanical oscillations. The nanowires were illuminated by normally incident CW laser light polarized parallel to the silicon nitride nanowires, with 1310 nm wavelength to match the plasmonic resonance, and the backward and forward scattered light intensity was monitored.

The structure's mechanical resonances at 1.53 and 1.98 MHz, corresponding to narrower and wider nanowire respectively, were identified as peaks in the power spectral density of voltage from the backscattering photodetector with a spectrum analyser. These spectra reveal the natural mechanical resonances of the nanowires due to the transduction of resonantly-enhanced thermal (Brownian) oscillation to light modulation.[35] Here we will characterize optomechanical bistability arising from the wider nanowire.

*Observation of optomechanical bistability*

Out-of-plane mechanical oscillation of the metamaterial nanowire pair was then driven by a piezoelectric actuator acting on the silicon frame that supports the metamaterial nanowires. The actuator is driven sinusoidally at frequency $f = \frac{\Omega}{2\pi}$ with a voltage $U = \frac{\widehat{U}}{2}(1 + \sin \Omega t)$, where $\widehat{U}$ is the peak-to-peak volage. Figure 2 shows the relative modulation amplitude of scattered light for $\widehat{U} = 0.6\ \text{V}_{\text{p-p}}$. A hysteresis loop forms at the nanowire's mechanical resonance when the structure is driven with increasing (black) and then decreasing (red) frequency [panels (a, b)]. The observed modulation of back scattered light at 2.004 MHz is 14% (1.0%) for increasing (decreasing) driving frequency, corresponding to an optical contrast $C = 14$ between the two states. Results for forward scattered light are similar [panel (b)].

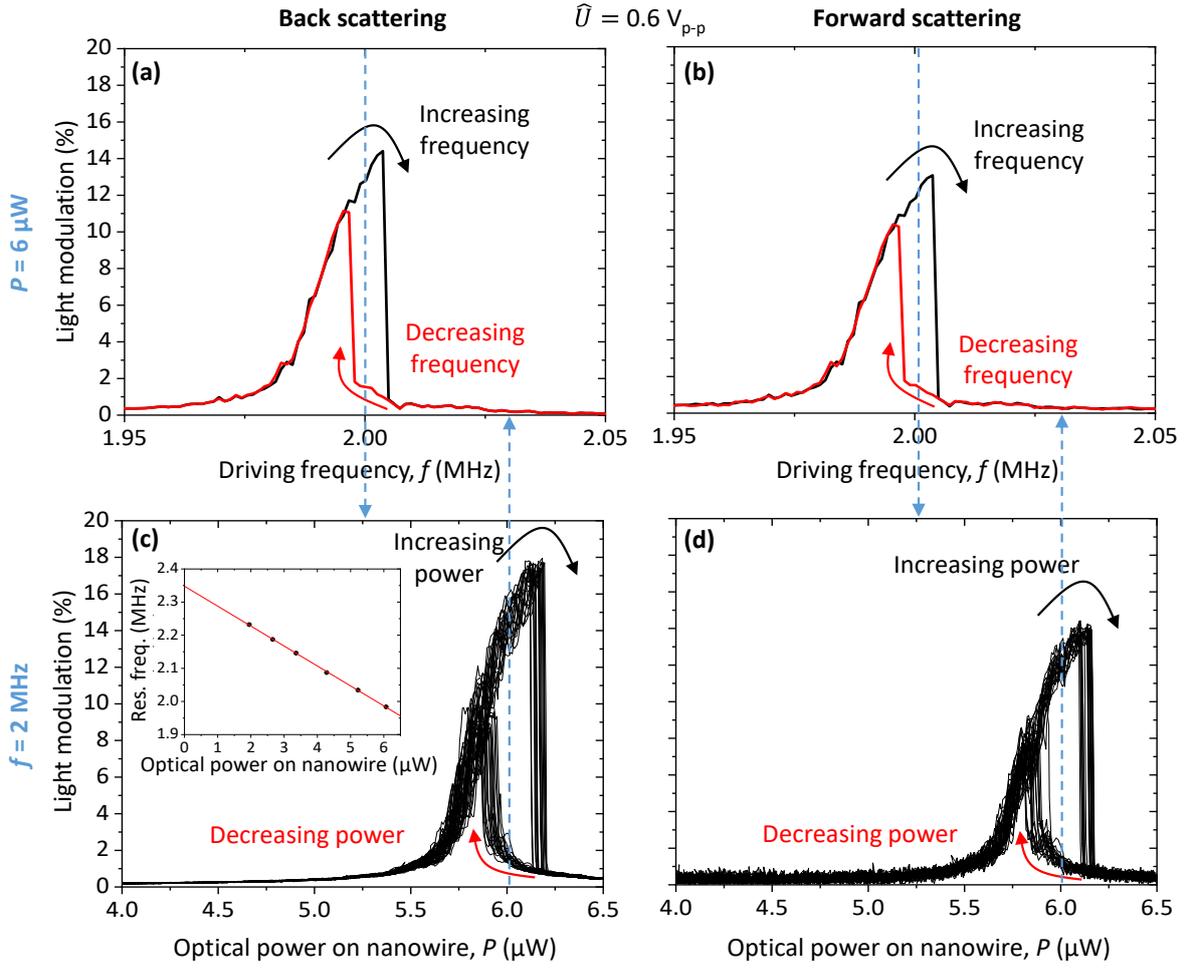

Fig. 2. Acoustically and optically controlled bistability. Modulation amplitude of (a,c) back and (b,d) forward scattered light for increasing (black) and decreasing (red) (a,b) acoustic excitation frequency and (c,d) optical power (18 cycles). The inset in (c) shows measurements of the mechanical resonance frequency as a function of incident optical power with a linear fit.

As the resonance frequency of the structure shifts to lower frequencies with increasing optical power due to stress reduction [panel (c) inset, Eq. (2)], increased optical power (at a fixed driving frequency) may be expected to trace the hysteresis cycle in the same direction as an increased driving frequency (at fixed optical power). To demonstrate optical control of the hysteresis cycle we keep the driving frequency fixed at 2 MHz (blue dashed lines) while the incident laser power is modulated with a period of $T = 1$s. Panels (c, d) show the measured light modulation amplitude for increasing (black) and decreasing (red) laser power. Indeed, a hysteresis loop forms in between 5.8 and 6.2 µW of optical power incident on the bistable nanowire.

*Tuning and dynamics of optical switching*

As the bistability arises from the mechanical nonlinearity of the metamaterial nanowire, a transition from linear to nonlinear and bistable oscillation may be expected with increasing driving amplitude. Figure 3a shows the amplitude of light modulation for increasing (black) and decreasing (red) driving frequencies $f$ and different amplitudes $\hat{U}$. At the lowest driving amplitude, the response of the wide nanomechanical beam is linear, the resonance profiles are symmetric and the modulation amplitude does not depend on the frequency sweep direction. With increasing driving amplitude, the resonance curves become increasingly asymmetric as mechanical nonlinearity of the system due to stiffening comes into play. The mechanical resonance shifts from its initial position at 1.978 MHz towards higher frequencies, reaching 2.004 MHz (1.995 MHz) for increasing (decreasing) driving frequency (Fig. 2a). The hysteresis loop forms and widens with increasing driving amplitude, showing that the plasmo-mechanical system translates its mechanical bistability into bistability of its optical properties.

The optical power required to engage the bistability can be controlled through the acoustic driving frequency (Fig. 2c inset). We observe that the optical bistability occurs at lower optical power for higher driving frequencies (Fig. 3b).

To study the dynamics of the bistability, we measure the hysteresis loop for different durations $T$ of the modulation cycle of the incident laser power (Fig. 3c). The hysteresis loop can be traced in $T = 100$ ms without any significant degradation. The contrast drops at shorter timescales, but a hysteresis loop is still observable in $T = 3$ ms. While conductive cooling estimates indicate a thermal response time of a few µs, a resonant system with quality factor $Q$ requires $Q$ cycles to respond to changes. With $Q \approx 300$ for the linear regime this corresponds to $\frac{Q}{f_0} \approx 150$ µs. The system needs to respond to many changes to trace the loop, which takes milliseconds.

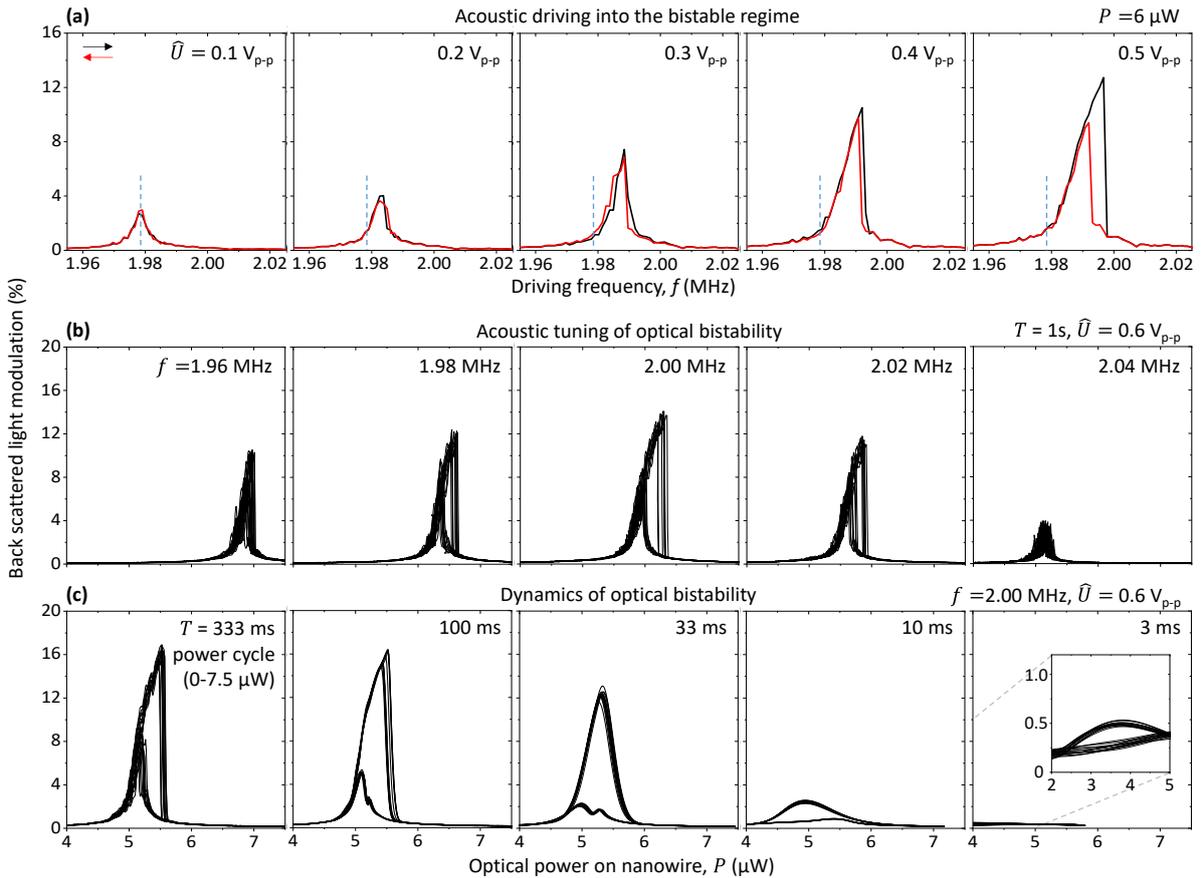

Fig. 3. Control over the optomechanical bistability. Modulation amplitude of light back scattered on metamaterial nanowires. (a) When the nanowires are driven to oscillate at increasing (black) and decreasing (red) frequencies with different driving amplitudes $\hat{U}$. (b,c) Hysteresis with increasing and decreasing optical power $P$ at different (b) acoustic driving frequencies $f$ and (c) durations $T$ of the laser power cycle.

## Discussion

At $f = 2$ MHz and $\hat{U} = 0.6$ V$_{\text{p-p}}$, the electrical power dissipated by the piezoelectric element is $Re\left(\frac{\langle U^2 \rangle}{Z^*}\right) = 42$ mW based on an impedance of $Z = (0.79 + i1.39)\Omega$, however, only a tiny fraction of this power is transferred to a nanowire, which is more than 11 orders of magnitude lighter than the piezoelectric element and the nanowire-supporting silicon frame. We calibrate the nanowire displacement based on the light fluctuations due to thermal oscillations at room temperature. We observe 0.05% RMS modulation of the collected back scattered radiation due to 198 pm thermal RMS displacement of the central part of the wider nanomechanical beam,[35] indicating that the largest detected light modulation of 14% for the driven nanowire arises from about $\hat{z} = 57$ nm peak-to-peak displacement. Considering a spring constant $k = m_{\text{eff}}\Omega_0^2 = 0.10$ N m$^{-1}$ based on the nanowire's effective mass of $m_{\text{eff}} = 0.65$ pg and the observed angular resonance frequency $\Omega_0 = 2\pi f_0$, this displacement corresponds to an energy of $E = \frac{1}{2}k\left(\frac{\hat{z}}{2}\right)^2 = 42$ aJ, and mechanical power dissipation of $\frac{\Omega_0 E}{Q} = 1.8$ pW based on a quality factor of $Q \approx 300$. The observed optical contrast $C = 14$ between the bistable states implies that the low-modulation state has $C$ times lower displacement and $C^2$ times lower mechanical energy.

Displacements of 10s of nanometres are expected to force the nanowire to oscillate in the nonlinear regime since the critical amplitude where nonlinearity comes into play is directly proportional to the structure's thickness, $a_{\text{cr}} \approx 1.46\, h/\sqrt{Q}$ [29, 36]. For the nanomechanical beam measured here, assuming 100 nm thickness, this translates to a critical peak-to-peak displacement of roughly 17 nm.

The optical energy required to switch between the bistable states of a nanowire may be estimated from the optical power and time required to switch. As shown in Fig. 3c, a hysteresis cycle with two switching operations can be achieved in $T = 100$ ms at optical power $P = 6\,\mu$W and the state switching energy can be evaluated as $\frac{PT}{2} = 300$ nJ.

Our results show that just a few µW of optical power are enough to switch between the two mechanical states of a nanowire, which is orders of magnitude lower than the power level requirements of integrated bistable systems based on nonlinear processes of silicon such as the thermo-optic effect[9] and carrier generation[37, 38] that usually require mW levels of input optical power, although thermal isolation of the ring structure can decrease the power required down to tens of µW[8]. Another way of introducing bistability in such integrated systems is through optically induced mechanical deformations. Optomechanical ring structures that exhibit bistability when driven by optical forces[10, 39] usually require input power levels in the hundreds of µW range. The lowest reported power levels have been achieved using high Q factor, low volume photonic crystal nanocavities[6]. Generally, such structures rely on high-Q optical resonances to achieve bistability at low power, implying that their operation is confined to a narrow wavelength range. In contrast, the bistability reported here offers large optical bandwidth. As the optical bistability arises from mechanical nonlinearity, it can be achieved with low Q factor optical resonators, which control the absorption spectrum of the nanowires. A direct consequence of this is that the structure can operate on a broad range of optical wavelengths around its absorption resonance.

# Conclusion

In conclusion, we have demonstrated that the oscillation of metamaterial nanowires can be driven into the nonlinear and bistable regime. In the bistable regime, the amplitude of resonant mechanical oscillation becomes dependent on both the structure's history of previous mechanical excitation and its history of previous optical illumination, and is observed as hysteretic modulation of the structure's optical properties. Thus, the mechanical bistability of nanowires yields optical bistability, i.e. optically-controlled switching between states with different optical properties that are stable under identical conditions.

A structure consisting of 2 nanomechanical metamaterial beams has been driven to oscillate in the highly nonlinear regime by a piezoelectric actuator achieving acoustically controlled hysteretic behaviour of its optical properties at pW levels of mechanical power supplied to a metamaterial nanowire. Optical switching between the bistable states is possible with only a few µW of optical power and the bistability is inherently optically broadband.

Our results indicate that nanomechanical metamaterials may offer all-optical solutions for switching and memory in photonic systems, as well as optical detection of increase/decrease of vibration frequencies (chirp) and optical power.

# Methods

**Sample fabrication.** The nanomechanical sample was fabricated by focused ion beam milling (FEI Helios 600 NanoLab) of a 50 nm-thick, low stress (<250 MPa) silicon nitride membrane (Norcada Inc.) that was covered with a 50 nm gold film by thermal evaporation. The metamaterial nanowires were fabricated on a 15.8 x 400 µm$^2$ rectangular membrane supported by a 5 x 5 mm$^2$ silicon frame of 0.2 mm thickness.

**Sample characterization.** Throughout all experiments and simulations, incident electromagnetic waves were linearly polarized with the electric field parallel to the silicon nitride nanowires. The experiment was conducted at a pressure of 4 x 10$^{-3}$ mbar. The power spectral density measurements were conducted with a normally incident CW laser beam at 1310 nm wavelength with a FWHM spot size of about 5 µm. Considering the unit cell, we consider effective nanowire widths of 425 and 375 nm for the calculation of the incident optical power per nanowire, which treats light incident on the unit cell as incident on the closest nanowire. The reflected laser beam was detected with a photodetector (New Focus Inc. 1811) and an electrical spectrum analyser (Zurich Instruments UHFLI). To mechanically excite the sample, the membrane supporting frame was placed on a piezoelectric actuator (Thorlabs) consisting of multiple PZT layers with a through-hole in its centre. Modulation of back and forward scattering on the sample were measured with a lock-in amplifier (Zurich Instruments UHFLI). Light modulation is given as relative peak-to-peak modulation in all cases.

**Acknowledgements.** This work is supported by the UK's Engineering and Physical Sciences Research Council (grants EP/M009122/1 and EP/T02643X/1) and Singapore Ministry of Education (grant MOE2016-T3-1-006 (S)).